\documentclass{article}
\usepackage{bibspacing}
\usepackage{spconf,amsmath,graphicx}
\usepackage{algorithm}
\usepackage{multirow}
\usepackage{algorithm}
\usepackage[noend]{algpseudocode}
\usepackage{amsfonts}
\usepackage{comment}
\usepackage{graphics}
\usepackage{caption}
\makeatletter
\newcommand{\thickhline}{%
    \noalign {\ifnum 0=`}\fi \hrule height 1pt
    \futurelet \reserved@a \@xhline
}

\title{Meta-Learning for Robust Child-Adult Classification from Speech}

\name{Nithin Rao Koluguri$^1$, Manoj Kumar$^1$, So Hyun Kim$^2$, Catherine Lord$^3$, Shrikanth Narayanan$^1$}
\address{
  $^1$Signal Analysis and Interpretation Laboratory, University of Southern California\\
  $^2$Center for Autism and the Developing Brain, Weill Cornell Medicine\\
  $^3$Semel Institute of Neuroscience and Human Behavior, University of California Los Angeles}
\begin{document}
\fontsize{9.3pt}{11.4pt}\selectfont
%
\maketitle
\begin{abstract}
Computational modeling of naturalistic conversations in clinical applications has seen growing interest in the past decade. An important use-case involves child-adult interactions within the autism diagnosis and intervention domain. In this paper, we address a specific sub-problem of speaker diarization, namely child-adult speaker classification in such dyadic conversations with specified roles. Training a speaker classification system robust to speaker and channel conditions is challenging due to inherent variability in the speech within children and the adult interlocutors. In this work, we propose the use of meta-learning, in particular prototypical networks which optimize a metric space across multiple tasks. By modeling every child-adult pair in the training set as a separate task during meta-training, we learn a representation with improved generalizability compared to conventional supervised learning.
We demonstrate improvements over state-of-the-art speaker embeddings (x-vectors) under two evaluation settings: weakly supervised classification (upto 14.53\% relative improvement in F1-scores) and clustering  (upto relative 9.66\% improvement in cluster purity). Our results show that protonets can potentially extract robust speaker embeddings for child-adult classification from speech.

\end{abstract}

\section{INTRODUCTION}
\label{sec:intro}

Automated child speech understanding using machines is an inherently more difficult problem than that of adult speech. A variety of factors have been identified and addressed in recent years, both from a signal processing viewpoint (such as large within- and across-age and gender variability due to a developing vocal tract \cite{lee1999acoustics}, errors in semantic structure of spoken language \cite{HAZAN2000377}), and limited data availability which has necessitated data augmentation and transfer learning techniques \cite{shivkumar_18arxiv}. 

An additional layer of complexity arises when addressing clinical and mental health applications involving children, where the condition may give rise to speech and language abnormalities. One such domain is autism spectrum disorder (ASD). ASD refers to a complex group of neuro-developmental disorders that are commonly characterized by social and communication idiosyncrasies, and whose reported prevalence in US children has been steadily rising (1 in 59 children as of 2014 \cite{baio2018prevalence}).
Child-adult interactions have been used in the ASD domain primarily for diagnosis (ADOS \cite{lord2000autism}) and measuring intervention response (BOSCC \cite{grzadzinski2016measuring}). 
Automated computational processing of the participants' audio \cite{bone2016use} and language streams \cite{kumar2016objective} has provided objective descriptions that characterize the session progress and understanding the relation with symptom severity.

However, behavioral feature extraction in above studies has necessitated manual annotation for speaker labels, which can be expensive and time-consuming to obtain especially for large corpora.
Automatic speaker label extraction involves a combination of speech activity detection (speech/non-speech classification) and speaker classification (categorization of speech regions into \textit{child} and \textit{adult}). In this work, we assume that oracle speech activity detection is available and focus on building a robust child-adult classification model.

Interest in child-adult speaker classification in spontaneous conversations has increased recently. 
Some of the early works used traditional feature representations such as MFCC, PLP and i-vectors, and clustering techniques such as Bayesian information criterion, information bottleneck and agglomerative hierarchical clustering \cite{wang2018progressive,cristia2018talker, zhou2016, sun2018, najafian2016speaker}. 
In \cite{wang2019centroid}, speech from children and adults was found to be sufficiently different in the embedding space to justify speaker classification.
A recent work using DNN embeddings (x-vectors) explored augmentation of child speech for PLDA training \cite{xiel2019}. The authors also observed that splitting the adult speech into gender-specific portions while training the PLDA returned improvements in diarization performance. However, similar to most of the above works, the authors do not make any distinctions within the child speech.

Training a conventional child-adult classifier from speech has at least two major issues: 
1) Large within-class variability especially for \textit{child} from age, gender, clinical symptom severity; and
2) Lack of sufficient amounts of balanced training data needed to tackle the above issue.
We propose to address the above issues using meta-learning, also known as learning-to-learn \cite{finn2017model}. Meta-learning consists of two optimizations: the conventional learner which learns within a task; and a meta-learner which learns across tasks. This is in contrast to conventional supervised learning, which operates within a single task for training and testing, and learns across samples. Meta-learning is inspired by the human learning process for rapid generalization to new tasks, for instance children who have never seen a new animal before can learn to identify them using only a handful of images. As a consequence, meta-learning has demonstrated success in low-resource applications \cite{snell2017prototypical, finn2017model} in computer vision in recent years.

In this work, we model each session in the training set as a separate task. Hence, each task consists of two classes: \textit{child} and \textit{adult} from the particular session. During training, classes are not shared across tasks, i.e., child in one session is a separate class from child in another session.
By optimizing the network to discriminate between child-adult speaker pairs across all training tasks (sessions), we mitigate the influence of within-class variabilities. Further, we remove the need for large amounts of training data by randomly sampling training and testing subsets (referred to as \textit{supports} and \textit{queries} respectively in meta-learning \cite{snell2017prototypical}) within each batch.
We evaluate our proposed method under two settings: 1) Weak supervision - a handful of labeled examples are available from the test session, and 2) Unsupervised - automated clustering. The latter is similar to conventional speaker clustering in diarization systems. We show that the learnt representations outperform baselines in both settings.

\vspace{-4mm}
\section{METHODS}
\label{sec:method}
\vspace{-2mm}
\subsection{Meta learning using prototypical networks}
\label{subsec:meta-learning}
\vspace{-1mm}
Meta-learning methods were introduced to address the problem of few-shot learning \cite{finn2017model}, where only a handful of labeled examples are available in new tasks not seen by the trained model. Deep metric-learning methods were developed within the meta-learning paradigm to specifically address generalization in the few-shot scenario. We choose prototypical networks (protonets) \cite{snell2017prototypical} which presume a simple learning bias when compared to other metric-learning methods, and have demonstrated state-of-the-art performance in image classification \cite{finn2017model} and natural language processing tasks such as sequence classification \cite{yu2018diverse}. 
Protonets learn a non-linear transformation into an embedding space where each class is represented using a single sample, specifically the centroid of examples from that class. During inference, a test sample is assigned to the nearest centroid.

Our application of protonets for speaker classification is motivated by the fact that participants in a test session represent unseen classes, i.e., speakers in an audio recording to be diarized are typically assumed unknown.
However, the target roles namely \textit{child} and \textit{adult} are shared across train and test sessions. Hence, by treating child-adult speaker classification in each train session as an independent task, we hypothesize that protonets learn the common discriminating characteristics between \textit{child} and \textit{adult} classes irrespective of local variabilities which might influence the task performance.

As a metric-learning method, protonets share similarities with triplet networks \cite{tristounet} and siamese networks \cite{chen2011learning} for learning speaker embeddings. Other than a recently proposed work which used protonet loss function for speaker ID and verification \cite{wang2019centroid}, to the best of our knowledge this work is one of the early applications of protonets for speaker clustering. Following, we illustrate the protonet training process using a single batch, then extend it to multiple training sessions.

\subsubsection{Batch training}
Consider a set of labeled training examples from $C$ classes $(\mathbf{X_{tr}},Y_{tr})$ = $\{(\mathbf{x_1}, y_1), ... (\mathbf{x_N}, y_N)\}$ where each sample $\mathbf{x_i}$ is a vector in $D$-dimensional space and $y_i \in \{1,2,..,C\}$. Protonets learn a non-linear mapping $f_{\theta} : \mathbb{R}^{D} \rightarrow \mathbb{R}^M$ where the prototype of each class is computed as follows:
\begin{equation}
\label{eqn:proto_basic}
    \mathbf{p}_{c} = \frac{1}{|S_{c}|} \sum_{(x_{i},y_{i})\epsilon S_{c}} f_{\theta}(\mathbf{x_{i}})
\end{equation}
$S_c$ represents the set of train samples belonging to class $c$. For every test sample $(x,y)$, the posterior probability given class $c$ is as follows:
\begin{equation}
\label{eqn:sofmax-eq}
    p_\theta(y=c | x)=\frac{\exp \left(-d_{\varphi}\left(f_{\theta}(\mathbf{x}), \mathbf{p}_{c}\right)\right)}{\sum_{c^{\prime} \in C} \exp \left(-d_{\varphi}\left(f_{\theta}(\mathbf{x}), \mathbf{p}_{c^{\prime}}\right)\right)}
\end{equation}
\noindent $d_\varphi$ denotes distance metric. While the choice of $d_\varphi$ can be arbitrary, it was shown in \cite{snell2017prototypical} that using Euclidean distance is equivalent to modeling the \textit{supports} using Gaussian mixture density functions, and empirically performed better than other functions. Thus, we use Euclidean distance in this work.
Learning proceeds by minimizing the negative log probability for the true class using gradient descent.
\begin{equation}
\label{eqn:proto_backprop}
 L(y,\mathbf{x}) = -\sum_{c=1}^{C} y_{c} \log(p_\theta(y=c \mid \mathbf{x})) 
\end{equation}
Pseudo-code for training a batch is provided in Algorithm \ref{alg:proto}.

\begin{algorithm}[t]
\caption{Single batch of protonet training}
\label{alg:proto}
\hspace*{\algorithmicindent} $\mathcal{U}$(S,N) denotes uniform sampling of $N$ elements from $S$ 
\hspace*{\algorithmicindent} without replacement \\
\hspace*{\algorithmicindent} \textbf{Input:} $D = \bigcup_{c=1}^{C} D_c$, where $D_c = \{(x_i,y_i) \; ; y_i = c\}$ \\
\hspace*{\algorithmicindent} \textbf{Output:} $L$ (Batch training loss) 
\begin{algorithmic}[1]
\For{$c$ in $\{1,...,C\}$} 
    \State $S_c = \mathcal{U}(D_c,k)$ \Comment{Supports}
    \State $Q_c = \mathcal{U}(D_c \setminus S_c, B-k)$ \Comment{Queries}
    \State $p_c = \frac{1}{k}\sum_{(x_i,y_i) \in S_c} f_\theta(x_i)$ \Comment {Prototypes}
\EndFor
\State $L \gets 0$
\For{$c$ in $\{1,...,C\}$} 
    \For{$(x,y)$ in $Q_c$}
        \State $L \gets L + \frac{1}{C\times(B-k)}[-\log(p_\theta(y=c|x)]$
    \EndFor
\EndFor
\end{algorithmic}
\end{algorithm}

\subsubsection{Extension to multiple sessions}
Consider $S$ sessions in the training corpus, with $N_{c,s}$ number of samples belonging to class $c$ in session $s$. We iterate through each session $s$, and randomly sample $k$ examples each from \textit{child} and \textit{adult} without replacement. These samples (\textit{supports}) are used to construct the prototypes using Equation (\ref{eqn:proto_basic}). From the remaining $N_{c,s}-k$ samples, $B-k$ samples are chosen without replacement from each class, where $B$ denotes the training batch size. These samples (\textit{queries}) are used to update the weights in a single back-propagation step according to Equation (\ref{eqn:proto_backprop}). Although a significant fraction of samples are not seen during a single epoch (1 epoch $\equiv$ $S$ batches), random sampling of supports and queries over multiple epochs improve the generalizability of protonets. 

\subsection{Siamese networks}
For unsupervised evaluation (clustering), we compare protonets with siamese networks \cite{koch2015siamese}, which learn a metric space to maximize pairwise similarity between same-class pairs and minimize similarity between different-class pairs. Specifically, we implement the variant used in speaker diarization \cite{garcia2017speaker}, where the training label for each input pair represents the probability of belonging to the same speaker. The network jointly learns both the embedding space and distance metric for computing similarity. In our work, we randomly select same-speaker (child-child, adult-adult) and different speaker (child-adult) x-vector pairs to provide input to the model. 

\begin{figure}[t!]
    \centering
    \includegraphics[width=0.45\textwidth]{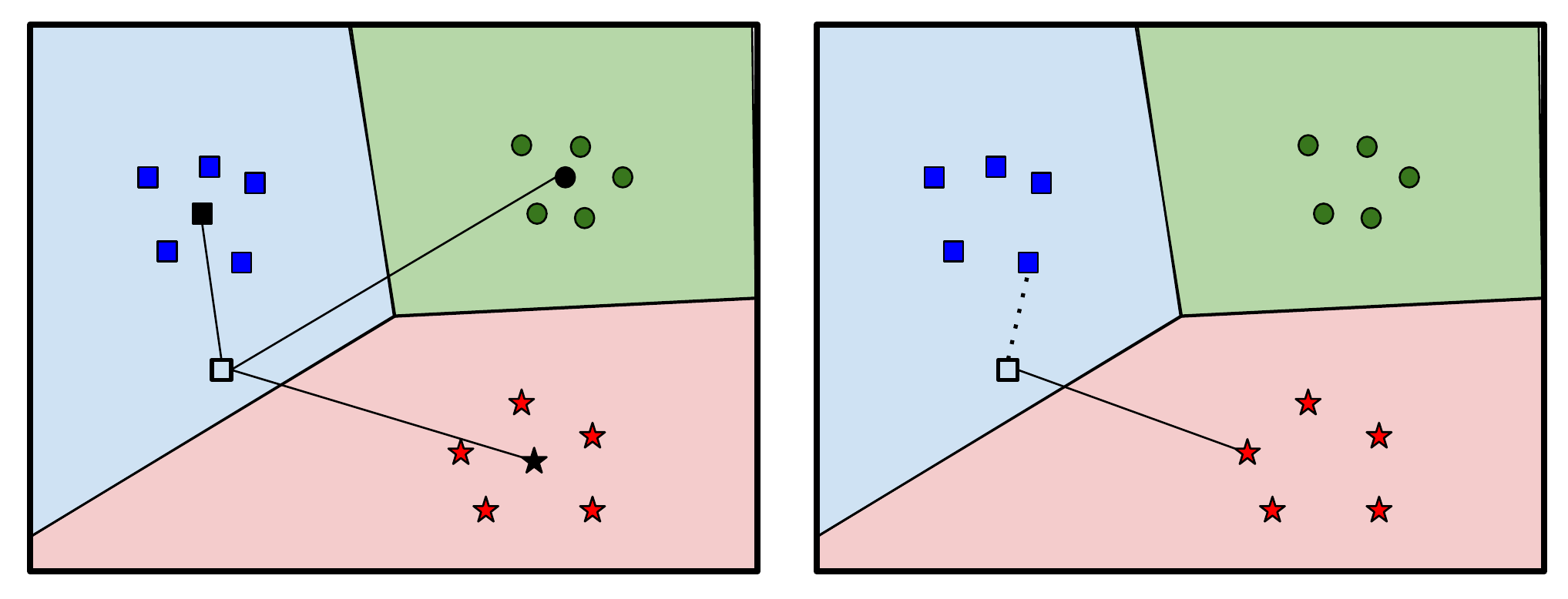}
    \caption{Training in protonets \textbf{(left)} vs siamese networks \textbf{(right)} in the embedding space. Colored backgrounds represent class decision regions. Distances from the query sample (non-filled) to prototypes from each class (filled with black) are used to estimate loss training loss using Equations (\ref{eqn:sofmax-eq}) and (\ref{eqn:proto_backprop}). siamese networks are trained to maximize similarity between same-speaker pairs (dashed line) and minimize similarity between different-speaker pairs (solid line). Illustration adopted from \cite{snell2017prototypical,wang2019centroid}.} 
    \label{fig:proto_vs_siamese}
\end{figure}

Fig. \ref{fig:proto_vs_siamese} illustrates the differences between siamese networks and protonets during training.

\begin{table}[t!]
\caption{Statistics of child-adult corpora used in this work.}
\label{tab:dataset}
\resizebox{0.48\textwidth}{!}{%
\begin{tabular}{ccccc} \thickhline
\multirow{2}{*}{\textbf{Corpus}} & \textbf{Duration}\textit{(min)} & \textbf{Child Age}\textit{(yrs)} & \multicolumn{2}{c}{\textbf{\# Utts}} \\
  & (mean $\pm$ std.) & (mean $\pm$ std.) &\textbf{Child} & \textbf{Adult} \\ \thickhline
ASD & 17.76 $\pm$ 11.99 & 9.02 $\pm$ 3.10 & 11045 & 20313\\ 
ASD-Infants & 10.35 $\pm$ 0.51 & 1.87 $\pm$ 0.78 & 1371 & 4120\\ \thickhline
\end{tabular}
}
\end{table}

\vspace{-4mm}
\section{EXPERIMENTS}
\label{sec:experiment}
\vspace{-2mm}
\subsection{Dataset}
\vspace{-1mm}
We select two types of child-adult interactions from the ASD domain: the gold-standard Autism Diagnostic Observation Schedule (ADOS \cite{lord2000autism}) which is used for diagnostic purposes and a recently proposed treatment outcome measure, Brief Observation of Social Communication Change (BOSCC \cite{grzadzinski2016measuring}) for verbal children who fluently used complex sentences. The ADOS Module 3 typically lasts between 45 and 60 minutes and includes over 10 semi-structured tasks. The ADOS produces a diagnostic algorithm score which can be used to classify children between ASD vs. non-ASD groups. On the other hand, BOSCC is a treatment outcome measure used to track changes in social-communication skills over the course of treatment in individuals with ASD, and is applicable in different collection settings (clinics, homes, research labs). A BOSCC session lasts typically for 12 minutes and consists of 4 segments (two 4-minute-play segments with toys and two 2-minute-conversation segments). We used a combination of ADOS (n=3) and BOSCC (n=24) sessions which were administered by clinicians and manually labeled by trained annotators for speaking times and transcripts. We refer to this corpus as \textit{ASD}. The sessions in \textit{ASD} cover sources of variability in child age, collection centers (4) and amount of available speech per child (Table \ref{tab:dataset}).

To check generalization performance, we train our models on \textit{ASD} and evaluate on a different child-adult corpus within the autism diagnosis and intervention domain. The \textit{ASD-Infants} corpus (Table \ref{tab:dataset}) consists of BOSCC (n=12) sessions with minimally verbal toddlers and preschoolers with limited language (nonverbal, single words or phrase speech). As opposed to \textit{ASD}, these sessions are administered by a caregiver, and represent a more naturalistic data collection setup aimed at early behavioral assessments with a familiar adult. The age differences between children in both corpora provides a significant domain mismatch.

\subsection{Features and Model Architecture}
\label{subsec:arch}
We use x-vectors from the CALLHOME recipe\footnote{https://kaldi-asr.org/models/m6} as pre-trained audio embeddings in this work, which have demonstrated state-of-the-art performance in speaker diarization \cite{SellSMGVMMDPWK18} and recognition systems \cite{snyder2018x}. X-vectors are fixed-length embeddings extracted from variable length utterances using a time-delay neural network followed by a statistics pooling layer. 
In all our experiments, 128 dimensional x-vectors are input to a feed-forward neural network with 3 hidden layers (128, 64 and 32 units per layer). Embeddings from the third hidden layer (32-dimensional) are treated as speaker representations. Rectified linear unit (ReLU) non-linearity is used in between the layers. Batch-normalization and dropout ($p$ = 0.2) are used for regularization. 
Adam optimizer ($lr$ = $3\mathrm{e}{-4}$, $\beta_1$ = 0.9, $\beta_2$ = 0.999) is used for weight updates. A batch size of 128 samples is employed.
Since \textit{ASD} corpus contains only 27 sessions, we use nine-fold cross validation to estimate test performance. 
At each fold, 18 sessions are used for model training. The best model is chosen using validation loss computed with 6 sessions. The remaining 3 sessions are treated as evaluation data. No two folds share the data from same speaker.

\subsection{Evaluation}

\subsubsection{Weak Supervision}

We evaluate our models in a few-shot setting similar to the original formulation of protonets \cite{snell2017prototypical} which is equivalent to sparsely labeled segments from the test session. 
In practice, such labels can be made available from the session through random selection or active learning \cite{settles2009active}.
We train a baseline model using the architecture from Section \ref{subsec:arch} and a softmax layer to minimize cross-entropy loss between \textit{child} and \textit{adult} classes. This model is directly used to estimate class posteriors on the testing data.
We refer to this model as \textit{Base}. We use a second baseline where the labeled samples from test sessions in each fold are made available during the training process, i.e., updating protonet weights using back-propagation \textit{(Base-backprop)}. 

For protonets, we train two variants: \textit{P20} and \textit{P30} with 20 and 30 \textit{supports} per class during training. A larger number of \textit{supports} translates into more samples for reliable prototype computation, however it results in fewer \textit{queries} for back-propagation.
During evaluation, 5 samples from each class in the test session are randomly chosen as training data. These samples are used to compute prototypes for \textit{child} and \textit{adult} followed by minimum-distance based assignment for the remaining samples in that session.
In order to estimate a robust performance measure for \textit{Base-backprop}, \textit{P20} and \textit{P30}, we repeat each evaluation 200 times by selecting a different set of 5 samples and compute the mean macro (unweighted) F1-score over the corpus. 

\subsubsection{Unsupervised: Clustering}
Clustering x-vectors using AHC and PLDA scores (trained with supervision) is an integral part of recent diarization systems \cite{SellSMGVMMDPWK18}. This method forms our first baseline. We note that the training data for PLDA transformation represents significant domain mismatch with our corpora. 
We use k-means and spectral clustering (using cosine-distance based affinity matrix) as unsupervised clustering methods for comparing x-vectors, siamese embeddings and protonet embeddings. In the siamese network, the distance measure between a segment pair is learnt between outputs from the third hidden layer (32-dimensional). For protonets, we use the models trained for weak supervision and extract embeddings at the prototype space (32-dimensional) for clustering. 
We use purity as the clustering metric, which describes to what extent samples from a cluster belong to the same speaker.

\begin{table}[hbt!]
\centering
\caption{Child-adult classification results using macro-F1 (\%)}
\label{tab:classification}
\begin{tabular}{ccc} 
\hline
\textbf{Method} & \textbf{ASD} & \textbf{ASD-Infants} \\ \hline
Base            & 82.67        & 53.67                \\
Base-backprop   & 78.64        & 56.29                \\
P20             & \textbf{86.66}        & 61.30                \\
P30             & 86.10        & \textbf{61.47}                \\ \hline
\end{tabular}
\end{table}

\vspace{-4mm}
\section{Results}
\vspace{-2mm}
\subsection{Classification}
\vspace{-1mm}
Weakly-supervised classification results are presented in Table \ref{tab:classification}.
In general, both variants of protonet outperform the baselines significantly in their respective corpora (ASD: $p$ \textless 0.05, ASD-Infants: $p$\textless 0.01). 
However, all models degrade in performance on the \textit{ASD-Infants} corpus as compared to \textit{ASD}. As mentioned before, the data from younger children presents a large domain mismatch between training and evaluation data and we suspect this as the primary reason for lower performance.
Surprisingly in \textit{ASD}, updating network weights using samples from test session (\textit{Base-backprop}) reduces classification performance. We suspect that the network overfits on the labeled samples.
However in the case of \textit{ASD-Infants}, the labeled samples from the test session provide useful information about the speakers resulting in modest improvement over a weaker \textit{Base}. While protonets provide the best F1-scores in both corpora, the performance in \textit{ASD-Infants} leaves room for improvement.
We do not observe any significant difference between \textit{P20} and \textit{P30}, suggesting that the performance is robust to the number of \textit{supports} and \textit{queries} during training.

\begin{table}[h!]
\centering
\caption{Speaker clustering results using purity (\%)}
\label{tab:clustering}
\begin{tabular}{ccccc} \hline
\multirow{2}{*}{\textbf{Method}} & \multicolumn{2}{c}{\textbf{ASD}} & \multicolumn{2}{c}{\textbf{ASD-Infants}} \\
 & \textbf{K-Means} & \textbf{SC} & \multicolumn{1}{c}{\textbf{K-Means}} & \multicolumn{1}{c}{\textbf{SC}} \\ \hline
x-vectors & 77.05 & 75.22 & 77.98 & 75.97 \\
siamese & 78.22 & 79.18 & 78.30 & 76.86 \\
P20 & \textbf{81.39} & \textbf{80.70} & \textbf{85.51} & \textbf{85.55} \\
P30 & 79.80 & 80.24 & 83.57 & 83.26 \\ \hline
\end{tabular}
\end{table}

\subsection{Clustering}
Clustering x-vectors using AHC and PLDA scores results in a purity of 63.45\% in \textit{ASD}, which is significantly lower than both K-means and Spectral Clustering (SC) for all the models in Table \ref{tab:clustering}. This suggests that the supervised PLDA models may be susceptible to unknown speaker types. Unsupervised PLDA adaptation using x-vectors' mean and variance from \textit{ASD} marginally improves the performance to 64.32\%, hence we do not include this method in the rest of our comparisons. As opposed to classification, clustering performance does not degrade in \textit{ASD-Infants}, suggesting that discriminative information between \textit{child} and \textit{adult} speakers within a session is preserved in all the embeddings compared in Table \ref{tab:clustering}. siamese networks present a modest improvement over x-vectors, upto 5.26\% relative improvement for spectral clustering in \textit{ASD}. However, protonets provide the best performance in both the corpora. In particular, \textit{P20} results in slightly higher purity scores than \textit{P30} across clustering methods and corpora. Hence, a larger number of \textit{queries} within a batch appears beneficial for speaker clustering in this work. We also note that the best clustering performance (\textit{P20}) is better in the out-of-domain corpus. We believe that the younger ages of children in \textit{ASD-Infants} over \textit{ASD} might benefit the clustering process.

\begin{figure}
    \centering
    \includegraphics[width=0.50\textwidth]{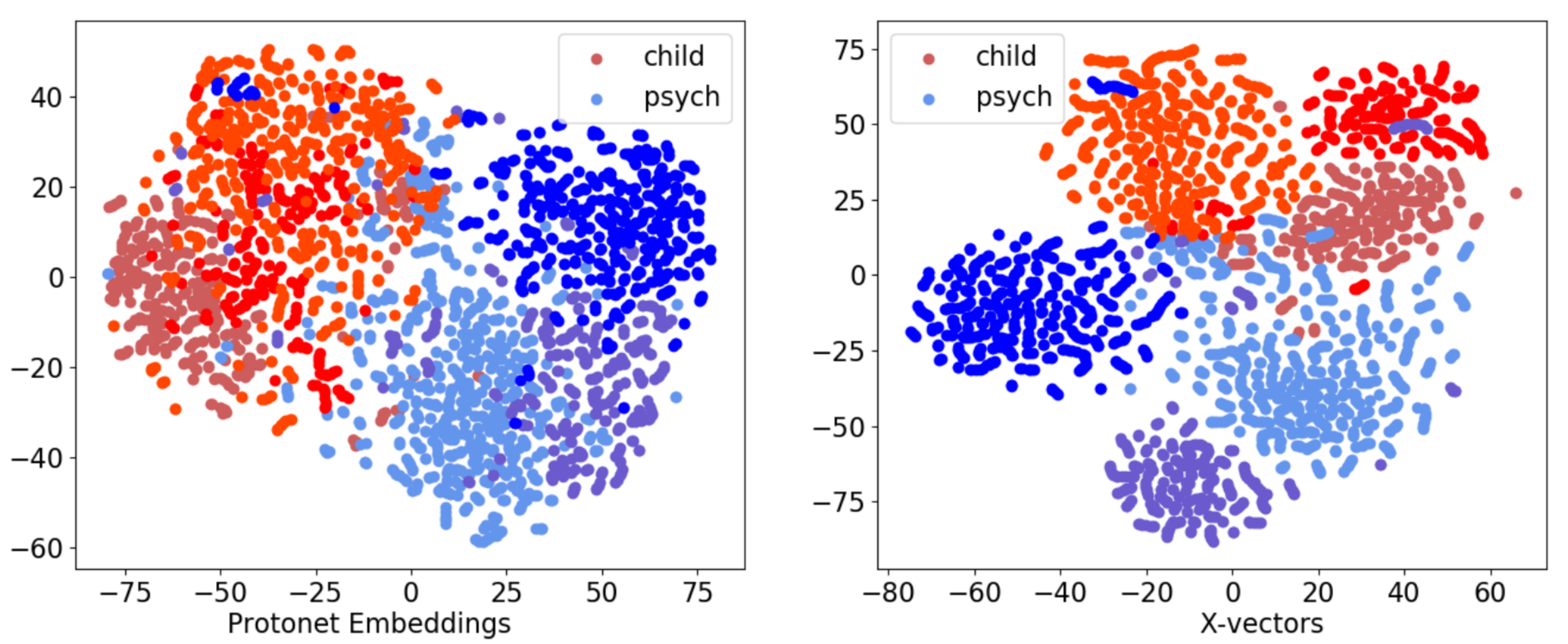}
    \caption{TSNE visualizations for protonet embeddings (\textbf{left}) and x-vectors (\textbf{right}) for 3 test sessions in \textit{ASD} corpora}
    \label{fig:tsne}
\end{figure}

\subsection{TSNE Analysis}
We provide a qualitative analysis using TSNE in Figure \ref{fig:tsne}. We collect embeddings from both \textit{child} and \textit{adult} from a single-fold (3 sessions) in \textit{ASD} and provide the TSNE visualizations for protonet embeddings and x-vectors. Embeddings from child and adult class are represented using 3 shades of red and blue respectively, one shade for each session. Although x-vectors cluster compactly within each speaker in a session, embeddings across sessions from the same class are spread apart. Protonets are able to cluster within classes compactly, while preserving the discriminative information between classes. In particular, embeddings belonging to \textit{child} (which are expected to cover more sources of variability) are as compact as embeddings from \textit{adult}. This suggests that protonets are able to learn across within-class variabilities for child-adult classification from speech.

\section{Conclusions}
In this work, we used meta-learning to perform child-adult speaker classification in spontaneous conversations. By modeling speaker classification from different sessions as separate tasks, we train protonets to learn speaker representations invariant to local variabilities. Using weakly-supervised and unsupervised settings, we show that protonets outperform x-vectors. Further, protonets outperform siamese networks for clustering when trained on the same input representations (x-vectors).
In the future, we would like to train a generic speaker diarization system using protonets. Protonets are a suitable choice for this problem, since an arbitrary number of speakers can be accommodated in every training session, and speaker identities need not be shared across sessions.  


\bibliographystyle{IEEEbib}
\bibliography{refs}

\end{document}